# PERFORMANCE ANALYSIS OF AN IMPROVED GRADED PRECISION LOCALIZATION ALGORITHM FOR WIRELESS SENSOR NETWORKS

Sanat Sarangi[1] and Subrat Kar[2]

[1,2]Bharti School of Telecommunication Technology and Management, IIT Delhi, Hauz Khas, New Delhi, India – 110 016

sanat.sarangi@gmail.com and subrat@ee.iitd.ac.in

### ABSTRACT

*In this paper an improved version of the graded precision localization algorithm GRADELOC, called IGRADELOC is proposed. The performance of GRADELOC is dependent on the regions formed by the overlapping radio ranges of the nodes of the underlying sensor network. A different region pattern could significantly alter the nature and precision of localization. In IGRADELOC, two improvements are suggested. Firstly, modifications are proposed in the radio range of the fixed-grid nodes, keeping in mind the actual radio range of commonly available nodes, to allow for routing through them. Routing is not addressed by GRADELOC, but is of prime importance to the deployment of any adhoc network, especially sensor networks. A theoretical model expressing the radio range in terms of the cell dimensions of the grid infrastructure is proposed, to help in carrying out a deployment plan which achieves the desirable precision of coarse-grained localization. Secondly, in GRADELOC it is observed that fine-grained localization does not achieve significant performance benefits over coarse-grained localization. In IGRADELOC, this factor is addressed with the introduction of a parameter that could be used to improve and fine-tune the precision of fine-grained localization..*

## 1. INTRODUCTION

The algorithm for graded precision localization (GRADELOC) proposed in [4] employs three localization principles: (a) centroid computation for coarse-grained localization (b) time difference of arrival based fine-grained localization and (c) last-mile localization with the help of a mobility module consisting of (i) an accelerometer for computing displacement based on steps and stride lengths and (ii) a gyroscope and magnetic-compass combination for estimating heading (direction) of motion. A node to be localized is referred to as NTL. There are three kinds of NTLs which need to be localized namely, coarse-grained NTL (CG-NTL), fine-grained NTL (FG-NTL) and extra-fine-grained NTL (EFG-NTL). CG-NTL only uses principle (a), FG-NTL uses (a) & (b) and EFG-NTL uses all three (a), (b) & (c) to achieve coarse-grained, fine-grained and extra-fine-grained localization respectively, with the same algorithm GRADELOC, thus justifying the phrase *graded precision localization*.

Localization with GRADELOC depends on two parameters: precision of location requirement level of the NTL and the transmission range of the grid nodes that constitute the sensor network infrastructure. Each grid cross-point has an energy unconstrained reference node set (REFN) consisting of two nodes: REFN0 for periodic beaconing and REFN1 for participation in TDOA based fine-grained localization. REFN0s periodically send beacons bearing their location to aid localization, whereas the REFN1s only come into play on receiving an explicit fine-grained localization request. They are assumed to participate in fine-grained localization on the principle outlined in [3]. The transmission range of the beaconing REFN0 nodes helps in segmenting the grid into regions each of which is covered by a different set of REFN0 nodes. As the NTL moves through the region, it keeps collecting beacons from REFN0 nodes. At periodic intervals (henceforth called centroid computation interval) it estimates its location by choosing the REFN0 nodes from which beacons meeting or exceeding a threshold are received and computes the centroid of their locations. For a CG-NTL this value is the advertised location. For FG-NTL, this value, if different from the last computed centroid, directs the FG-NTL to perform a fine-grained localization of itself with the help of REFN1 nodes. As compared to an FG-NTL, an EFG-NTL has the following additional functionality- it uses its mobility module to continuously estimate its location between two fine-grained localization operations.

It may be observed that the range of a REFN0 node in GRADELOC as shown in Fig. 3 does not cover any other neighbouring REFN0 nodes. Thus, the location estimated by the mobile node cannot be communicated through the REFN0 nodes to a central gateway placed (say) at one corner of the grid. Moreover, the regions formed in a grid-cell also indicate that the candidate REFN0 set (consisting of REFN0s from which beacons greater than or equal to a threshold value are received) keeps changing at a steady rate. So, performing a fine-grained localization in such conditions does not bring about a significant cost-performance trade-off where the cost is the communication overhead in the form of fine-grained localization requests and performance represents the improved precision of localization. Keeping these issues in mind, we suggest some improvements to the GRADELOC algorithm in the form of IGRADELOC.

## 2. THE IGRADELOC ALGORITHM

The flow diagram of IGRADELOC is as given in Fig. 2. IGRADELOC retains the same three NTL variants: CG-NTL, FG-NTL, and EFG-NTL with their parameter set *(coarseGrained, fineGrained, selfLocalize)* initialized as per Tab. 1.

## 3. TOPOLOGY

The topology used by IGRADELOC (same as that of GRADELOC) is a grid of 5x5 fixed nodes where each node is a (REFN0, REFN1) set, shown in Fig. 1. However the transmission range of each REFN0 is enhanced as given by Fig. 4. This enhanced range, when compared to that in Fig. 3 of GRADELOC, is able to handle routing through the grid nodes since neighbouring nodes are within the range of each other. The range of a REFN0 node is denoted by R and the side length of a square grid-cell is denoted by L. Normally, $R \geq L$ is a necessary condition to achieve routing, however, we discuss in Sec. 4 why it may not be a sufficient condition for successful operation of IGRADELOC.

Table 1: Node Classification

| NTL | coarseGrained | fineGrained | selfLocalize |
|---|---|---|---|
| None | False | False | False |
| CG-NTL | True | False | False |
| FG-NTL | True/ False | True | False |
| EFG-NTL | True/ False | True | True |

## 4. THEORETICAL FOUNDATIONS

Here we establish the basis on which we select the optimum values for various parameters under which IGRADELOC operates. The following sections attempt to address some problems in creating a deployment scenario for IGRADELOC. We assume that:

1. Fine-grained localization with TDOA works well when the speed of the NTL is low [3]. Since we are comparing the performance of different localization schemes here, the NTL is assumed to move at a constant low speed according to the mobility model discussed in Sec. 8.3.

2. The REFN0s and the NTLs are meant to be realized with the same sensor nodes. So, all discussions with respect to node characteristics e.g. radio range, apply to both REFN0s and NTLs.

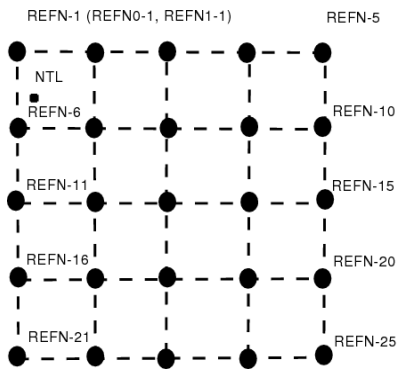

Figure 1: 5x5 REFNs arranged in a grid

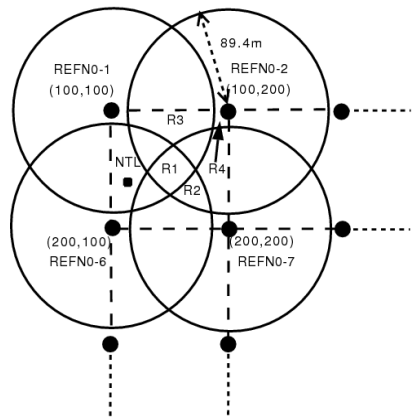

Figure 3: GRADELOC Radio Range

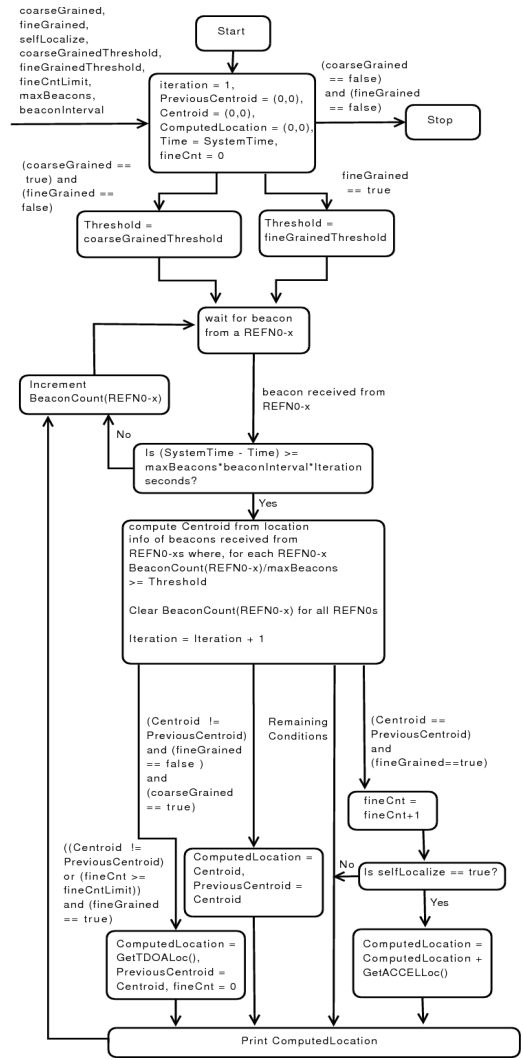

Figure 2: Flow diagram of IGRADELOC algorithm

152

## 5. RELATIONSHIP OF DISTANCE BETWEEN GRID NODES AND NODE RANGE

When an FG-NTL decides to do a fine-grained localization (FGL) of itself, it has to reach at least three non-collinear REFN1 nodes. We assert that the transmission range of the NTL moving within the fixed-grid is at least $L\sqrt5/2$ for it to have at least three non-collinear fixed nodes at one hop distance.

**Proof:** Consider grid-cell ABCD (with side of length L) of the fixed-grid in Fig 4 shown in Fig. 5. AC and BD are diagonals and EG and FH are perpendicular bisectors of the opposite sides dividing the square into eight congruent triangles P1,...,P8.

1. Let $p_i$ and $p_j$ be two points in ABCD. EG and FH are the perpendicular bisectors of the opposite sides and I is the center of the square.

2. Let S be a set of points $\{p_{1\ldots}\ p_N\}$ and $p_i, p_j \in$ S. $D(p_i, p_j)$ is the distance between $p_i$ and $p_j$. $M(p_i, p_j) = \max\{D(p_k, p_j)\}$ for all $p_k \in$ S, $1 \le k \le N$.

3. Let X = {All points in $\Delta$ AEI}.

4. For all $p_i \in$ X, clearly, $M(p_i, p_A) = D(p_I, p_A)$ (hypotenuse) = $L\sqrt5/2$ denoted by $L_1$.

5. Let $Y_1$ = {All points in $\Delta$ BAE}. For all $p_i \in Y_1$, clearly, $M(p_i, p_B) = D(p_E, p_B)$ (hypotenuse) = $L\sqrt5/2$ (denoted by $L_2$). Let $Y_2$ = {All points in $\Delta$ BGE}, clearly, $M(p_i, p_B) = D(p_E, p_B)$ (hypotenuse). Since, $X \subset Y_1 \cup Y_2$, for all $p_i \in$ X $\cup \{p_B\}$, $M(p_i, p_B) = D(p_E, p_B)$ denoted by $L_2$.

6. Let Z = {All points in $\Delta$ AID} and for all $p_i \in$ Z, $M(p_i, p_D) = D(p_A, p_D)$ (hypotenuse). Since, $X \subset$ Z, for all $p_i \in$ X $\cup \{p_D\}$, $M(p_i, p_D) = D(p_A, p_D)$ (hypotenuse) = L, denoted by $L_3$.

From steps 4, 5 and 6 we can conclude that the three fixed REFN0 nodes at A, B and D are at one hop distance from an NTL with position $p_i$ where $p_i \in$ X provided the range of the NTL is $\ge \max(L_1, L_2, L_3) = L_2 = L\sqrt5/2$. By symmetry, it can be shown that the relation holds for each of the $\Delta$ s P1,...,P8. Hence, there are (at least) three REFN0 nodes that are at one hop distance from an NTL moving within a grid-cell provided, range of the NTL $\ge L\sqrt5/2$.

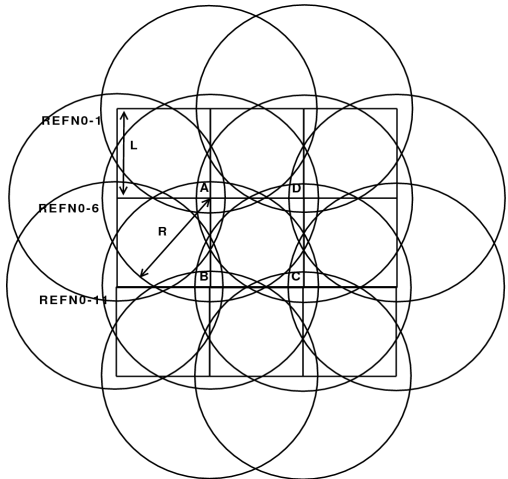

Figure 4: IGRADELOC Radio Range

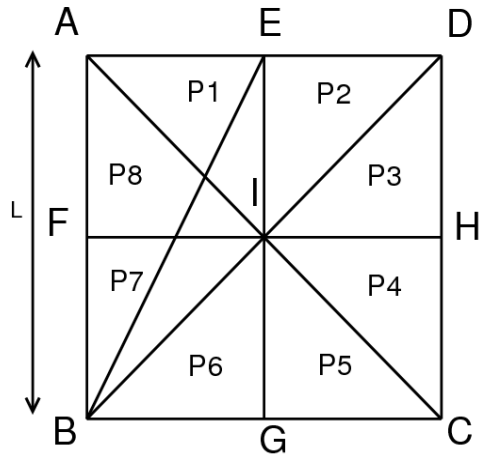

Figure 5: Analytical model of grid-cell ABCD for estimating the desirable transmission range for NTL

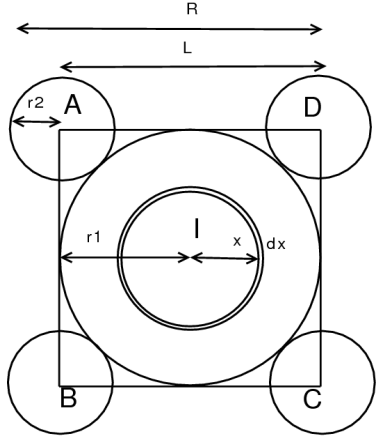

Figure 6: Analytical model of grid-cell ABCD for estimating localization error

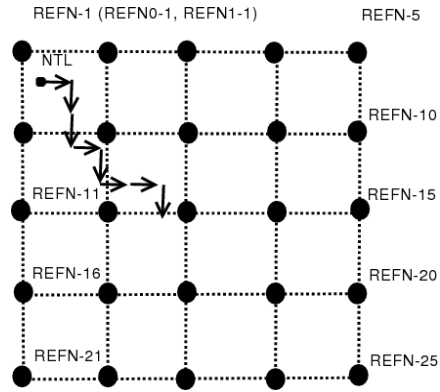

Figure 7: NTL mobility model

## 6. ANALYTICAL ERROR ESTIMATION MODEL FOR COARSE-GRAINED LOCALIZATION

Fig. 4 shows a typical deployment scenario for IGRADELOC where range of each REFN0 and NTL is $L\sqrt{5}/2$ as per Sec. 5 and the length of each square cell's side is L. For analysis we approximate the cell ABCD in Fig. 4 by the regions shown in Fig. 6. We denote a circle with center K and radius L as $C_{K,L}$. In Fig 6, radius of $C_{I,r1} = L/2$ and of $C_{A,r2}$, $C_{B,r2}$, $C_{C,r2}$ and $C_{D,r2} = (R - L)$. Let $S_E$ denote set of N estimated locations $\{E_1...E_N\}$ of an NTL (obtained by computing the centroid of the received beacons in case of coarse-grained localization), and $S_A = \{A_1...A_N\}$ be the corresponding set of actual locations. The cumulative localization error (CLE) and mean absolute error (MAE) of CLE are defined as follows:

$$CLE = \sum_{i=1}^{i=N} |E_i - A_i| \qquad (1)$$

$$MAE(CLE) = CLE/N \qquad (2)$$

Computing localization error for an NTL present anywhere in the four quarter circles at each corner inside grid-cell ABCD is equivalent to computing it for one of the circles, say $C_{A,r2}$. Areas of $C_{I,r1}$ and $C_{A,r2}$ together represent about 82% of the area of the square so estimating coarse-grained localization error in these regions can be used to approximately determine the coarse-grained localization error over the whole grid.

**Region 1:** Centroid of ABCD is I. The estimated location $E_i$ (computed by receiving beacons from grid nodes A, B, C, D) for an NTL at any point $p_i \in C_{I,r1} = I$ (centre). So, for an NTL present at a point $p_i$ on circumference of circle $C_{I,x}$, CLE = (circumference of $C_{I,x}$) x (radius of $C_{I,x}$) = $2\pi x.x$. Integrating over $C_{I,r1}$, cumulative localization error for an NTL present anywhere inside $C_{I,r1}$, denoted by $CLE_1$ is given by

$$CLE_1 = \int_0^{\frac{r1}{}} 2\pi x^2 dx = 2\pi (r1)^3/3 \qquad (3)$$

where r1 = L/2 and number of points denoted by $N_1$ is nothing but the area of $C_{L,r1} = \pi (r1)^2$

**Region 2:** Similarly, the cumulative localization error for an NTL present anywhere in $C_{A,r2}$ denoted by $CLE_2$ is given by

$$CLE_2 = \int_0^2 2\pi x^2 dx = 2\pi (r2)^3/3 \qquad (4)$$

where $r2 = (R-L)$ and corresponding number of points denoted $N_2$ is the area of $C_{A,r2} = \pi (r2)^2$. Putting the lower bound of the desirable range $R = L\sqrt{5}/2$ of an NTL, we get

$$MAE = (CLE_1 + CLE_2)/(N_1 + N_2) = 0.32L \text{ (approx.)} \qquad (5)$$

# 7. SELECTING BEACONS AND CENTROID COMPUTATION INTERVALS

Let the centroid be computed every P sec. and beacon interval be p sec. A granularity function $G(p, P) = p/P$ is defined with Threshold (T) for an NTL to select a REFN0 as candidate for centroid computation given by $T = G(p, P).n$, where $n \in$ Set of positive integers s.t. $0 < np/P \leq 1$. We use Fig. 6 to analyse the values to be assumed by the various parameters since they are homogenous (circular) spaces which form significant portions of the mobility area (Sec. 6) and the centroid computed by an NTL is nothing but the center of the respective circles. To ensure invocation of a centroid computation within these spaces, centroid computation interval P is taken as:

$$P = \min\{r1, r2\}/S = (\sqrt{5}/2 - 1)L/S \qquad (6)$$

where S is the maximum speed of the NTL. Accordingly, p is chosen so that $G(p,P) = 0.1$ and that $1 \leq n \leq 10$. The parameter *maxBeacons* in Fig. 2 refers to maximum value that could be taken by n. So *maxBeacons* has a value of 10. Parameter *beaconInterval* is nothing but p.

## 7.1. Fine-grained localization with TDOA

An FG-NTL or EFG-NTL invokes GetTDOALoc() (see Fig. 2) for a fine-grained localization of itself. To start the fine-grained localization, the NTL broadcasts a beacon giving the current location of the NTL which alerts at least three REFN1s (shown in Sec. 5) that an FGL is being demanded. [3] discusses the results of TDOA based localization using Taylor-series and Fang's algorithms for LOS and NLOS conditions which are used to solve the hyperbolic equations obtained during fine-grained localization with TDOA. For analysis, [3] uses three anchor nodes placed in a right angled triangle and a slow-moving mobile node within the triangle formed by the anchor nodes. Since this architecture is similar to that of GRADELOC and IGRADELOC, the results obtained in [3] for line-of-sight (LOS) conditions are used to define the (approximate) localization error that could be expected for the fine-grained localization operation. So, for the purpose of simulation, the GetTDOALoc() function in our algorithm essentially returns the actual location with a certain error to account for the localization error. In IGRADELOC, we introduce a parameter *fineCntLimit* which could be used to force an out-of-turn fine-grained localization if the computed centroid does not change for *fineCntLimit* centroid-computation steps. *fineCntLimit* is set using the NTL movement characteristics and the maximum approximate duration for which centroid may not be expected to change. If we consider Fig. 6, an NTL moving across $C_{L,r1}$ with a maximum speed of S, the computed centroid will not change for a distance 2(r1). Hence, for a centroid-computation-interval of P, $0 < fineCntLimit \leq \lfloor 2(r1)/(PS) \rfloor$ is used, provided $2(r1)/(PS) \geq 1$. To prevent an out-of-turn localization, an unattainably large value of *fineCntLimit* is chosen.

### 7.2. Self-localization with mobility module

We may calculate displacement accurately as in [2] and [1]. However, the displacement cannot be used for localization if the initial position error is not removed. Hence, for the results of GetACCELLoc() to be meaningful, it is required that an EFG-NTL has both *fineGrained* and *selfLocalize* enabled as shown in Tab. 1. In [1], the authors report the results of an experimental Pedestrian Navigation System, where they use accelerometer based step and stride information, a gyroscope & magnetic-compass combination for heading determination. We adopt a similar model shown in Fig. 7 where the EFG-NTL is assumed to be a pedestrian fitted with the module consisting of an accelerometer, gyroscope and magnetic compass. For analysis, we assume that all the three sensors are error prone. The accelerometer reports the stride length as a percentage SA of the actual length and step detection itself is governed by a percentage DA. Since our deployment scenario is a square grid, the NTL starts at the top-left corner and moves towards the bottom-right corner of the grid by randomly choosing a direction (right or down) after every N steps it takes in a particular direction. A heading error of $\theta°$ is introduced downwards (or to the right) for NTL's movement to the right (or downwards).

## 8. SIMULATION

A discrete-event simulation of the proposed algorithm is carried out with the parameters given in the following sub-sections. The behaviour of an NTL is analyzed by configuring it as a CG-NTL, FG-NTL and EFG-NTL by setting the parameters as per Tab. 1.

Table 2: Mapping of Error Index to Range

| Error Index | e (Absolute Error in m) |
|:-:|:-:|
| 1 | $0 \le e \le 2$ |
| 2 | $0 \le e \le 5$ |
| 3 | $0 \le e \le 10$ |
| 4 | $0 \le e \le 20$ |
| 5 | $0 \le e \le 30$ |
| 6 | $0 \le e \le 50$ |
| 7 | $0 \le e \le 75$ |

Table 3: NTL Type

| Index | NTL Type |
|:-:|:-:|
| 1 | CG-NTL |
| 2 | FG-NTL-Improved |
| 3 | FG-NTL |
| 4 | EFG-NTL-Accurate |
| 5 | EFG-NTL-Inaccurate |

### 8.1. Topology, Beaconing and radio parameters

A set of 25 nodes are arranged on a fixed-grid in a 300m x 300m field with L set to 75m which makes our desirable node range $\ge$ 84m, a realistic assumption for commonly available nodes. Taking L = 75m and S = 1s, Eqn. 6 gives us 8.85 as the value of P. A (close) value 10 is taken for P. Beacon interval p is 1 to achieve a granularity of 0.1. Threshold T is set to 0.9 (for both coarse-grained and finegrained). Each node has a transmission range of 84m. The Radio data-rate is 250 kbps with PSK as the modulation mechanism. Interference is used for collision detection at the receiver.

### 8.2. Parameters for fine-grained localization of (E)FG-NTL

As outlined in Sec. 7.1 Taylor-series method and Fang's method under LOS conditions result in errors in the range of (1m - 5.5m) and (1m - 7m) respectively. The result from Fang's method is used to estimate the error in fine-grained localization since it gives us a higher upper bound of the localization error. Essentially, the location (x,y) computed by GetTDOALoc() has an error given by ±(xerror, yerror) added to the actual location, where both the parameters are outcomes of two independent uniform pdfs with (qmin, qmax) as limits. Choosing ± (1, 5) as the value for (qmin, qmax) gives us the minimum and maximum MAE of $1\sqrt{2}$m( 1.4m) and $5\sqrt{2}$m( 7m). Since S is 1m/s , P = 10 and r1 = L/2 = 37.5m, *fineCntLimit* is taken as 4 to stay within the bounds of the limits discussed in Sec. 7.1. This improved version of FG-NTL is henceforth referred to as FG-NTL-Improved. For a normal FG-NTL, *fineCntLimit* is typically set to 100 (a large value) to prevent an out-of-turn fine-grained localization.

### 8.3. A step, stride based NTL mobility model

Two sets of values are used for the mobility model. A set of values SA = 95%, DA = 99%, GA = 5° represents a highly accurate equipment where, SA = 95% indicates that the reported stride length is 95% of the actual stride length, reported number of steps (DA) is 99% of the actual number and a heading error (GA) of 5° occurs on the lines of the results shown in [2]. The EFG-NTL with these parameters is henceforth called EFG-NTL-Accurate. We also use set SA = 90%, DA = 90%, GA = 10° to observe the degradation in performance, with inaccurate hardware and the corresponding EFG-NTL is called EFG-NTL-Inaccurate. For all purposes, the NTL takes one step per sec. Stride length SL (in m) of the step is determined by a uniformly distributed random variable in the range [0.7, 0.8].

## 9. RESULTS

### 9.1. Comparison of theoretical MAE and MAE obtained from simulation

The MAE for coarse-grained localization obtained from simulation as shown in Fig. 8 is found to agree with the theoretical error given by Eqn. 5 indicating that our analysis can be used for carrying out a pre-deployment network planning to determine the side length of each square grid-cell and the desirable radio range (which could be adjusted by possibly altering the transmission power level of the sensor node) to achieve the desired precision of coarse-grained localization.

### 9.2. Performance comparison of NTLs

Fig. 9 uses error index from Tab. 2 to show the error distribution for CG-NTL, FG-NTL, FG-NTL-Improved, EFG-NTL-Accurate and EFG-NTL-Inaccurate respectively. It is observed that EFG-NTL-Accurate has the best performance as expected with all errors within 10m (Error Index = 3) where as  CG-NTL has the poorest performance with less than 2% errors within 10m. EFG-NTL-Inaccurate has 89% errors lying within 10m. A notable performance improvement is seen with FG-NTL-Improved with 59% errors contained within 10m  when compared to FG-NTL for which it is 47%. The observed mean absolute error (MAE) and the root mean squared error (RMSE) for all the NTLs are shown in Fig. 10. The interpretation of NTL Type is given in Tab. 3. FG-NTL-Improved has lower MAE and RMSE values than FG-NTL with approximately 8% additional communication overhead.

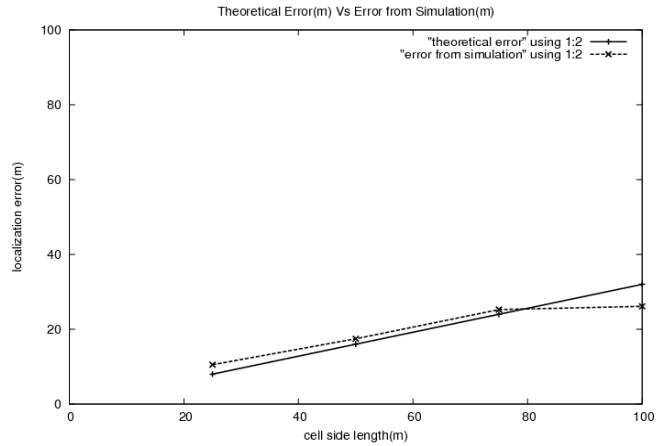

Figure 8: A comparison of theoretical values for
coarse-grained localization error and the
corresponding values obtained from simulation

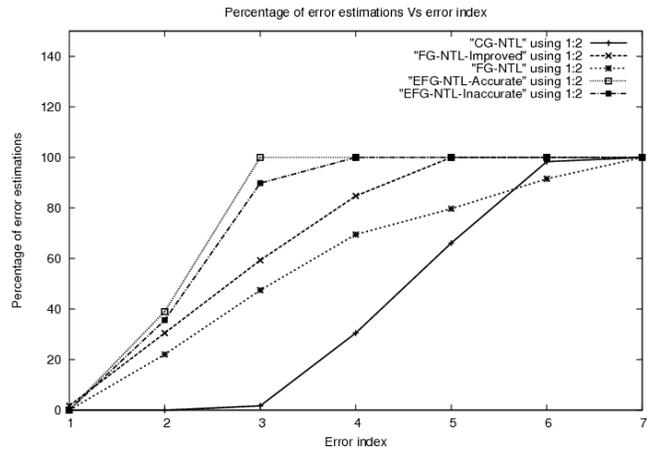

Figure 9: Distribution of localization errors
for various NTLs

## 10. ENERGY CONSIDERATIONS

With IGRADELOC, apart from graded expenditure of energy through design modularity, fine-grained localization initiation by notification of at least three REFN1s with a single broadcast from an NTL, and introduction of the *fineCntLimit* parameter for fine-tuning (localization) precision and cost (of communication) helps in further regulating the energy consumption. Each NTL spends energy according to its level of sophistication. CG-NTLs are just passive receivers. FG-NTL spend some additional energy for notifying REFN1s for a fine-grained localization, and EFG-NTLs spend some energy (in addition to the amount spent by FG-NTL) on their mobility module.

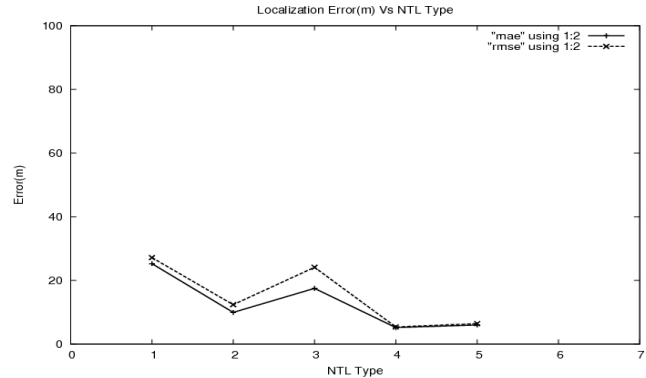

Figure 10: MAE, RMSE of localization errors for various NTLs

## 11. CONCLUSION

In IGRADELOC emphasis has been on improving the performance of FG-NTL and EFG-NTL within a deployment architecture justifying their need. The characteristics of the topology with respect to grid-cell dimensions and REFN0 & NTL transmission range have been analysed to select optimum configuration to achieve the desired performance. Further, the architecture now also provides for routing capability through the grid nodes.

**Authors**

Mr. Sanat Sarangi is a Ph.D. student at Bharti School of Telecommunication Technology and Management, IIT Delhi, India. His research interests include Embedded Systems, Computer Networks and Sensor Networks.

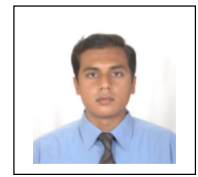

Dr. Subrat Kar is a professor in the Department of Electrical Engineering and Bharti School of Telecommunication Technology and Management, IIT Delhi. He received his Ph.D. from IISc. Bangalore, India. His research interests include Photonic Switching, Optical and Computer Communication Networks.

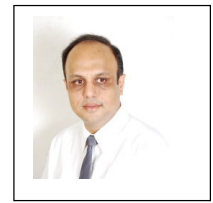